\documentclass[twoside,fleqn]{ActaStyle}
\usepackage{times,cite}

\usepackage{graphicx,epsfig}    
\usepackage[tbtags]{amsmath}    

\def\bea {\begin{eqnarray}}
\def\eea {\end{eqnarray}}
\def\be {\begin{equation}}
\def\ee {\end{equation}}
\def\nn {\nonumber}

\def\authorheading{Jajati K. Nayak et al.}
\def\shorttitle{Kaon to Pion ratio in Heavy Ion Collisions }

\newcommand{\pathnow}{}
\begin{document}
\pagerange{1}{4}

\title{Kaon to Pion ratio in Heavy Ion Collisions}

\author{
Jajati K. Nayak$^*$\email{jajati-quark@veccal.ernet.in *}, 
Jan-e Alam$^1$, Pradip Roy$^2$, Abhee K.Dutt-Mazumder$^2$\\
Bedangadas Mohanty$^1$
}
{
$^1$
Variable Energy Cyclotron Centre,Kolkata-64,India,\\
$^2$
Saha Institute of Nuclear Physics,Kolkata-64,India.
}

\date{\today}

\abstract{%
The momentum integrated Boltzmann equation has been used 
to study the evolution of strangeness of the strongly 
interacting system formed after the heavy ion collisions
at relativistic energies.We argue that the experimentally
observed non-monotonic, horn-like structure in the variation 
of the $K^+/\pi^+$ with colliding energy appears due to
the release of large number of colour degrees of freedom. 
}

Calculations based on lattice QCD \cite{lattice}
indicate a phase transition from  hadronic matter 
to a deconfined state of quarks and gluons, called quark gluon plasma
(QGP),  at 
a temperature, $T_c\sim 175$ MeV.
The quest for creating QGP in laboratory has  led to
heavy ion collision experiments being carried out at centre 
of mass energy ($\sqrt{s_{NN}}$) ranging from a few 
GeV to 200 GeV per nucleon. In the experimental side judicious choices 
of beam  energy, colliding ions, impact parameter and other 
suitable variables have been used to  confirm the existence 
of QGP~\cite{npa757}. 

The non-monotonic behaviour of  $K^{+}/\pi^+$ 
with colliding energy of the nuclei~\cite{MG} has led to 
intense theoretical activities~\cite{CORS,GG,BT,LR}. 
In the present work, we study the evolution of the strangeness
of the system formed in heavy ion collision
for various initial conditions within the frame work of momentum
integrated Boltzmann equation. 

In the QGP phase the evolution equation of strangeness is given by:
\begin{equation} 
\label{eq1}
\frac{dr_{\bar{s}}}{d\tau} = \frac{R_{\bar{s}}(T)}{n^{eq}_{\bar{s}}}
[ 1 - r_{s}r_{\bar{s}}]
\end{equation}
where, $r_i=n_i/n^{eq}_i$,
$n_i$ and $n^{eq}_i$ are the non-equilibrium and
equilibrium densities of the species $i$ in 
the QGP phase. $R_i$ is the rate of production
of particle $i$ at temperature $T$ and $\tau$ is the proper time.  
The reactions considered for the production of $\bar{s}$ are:
$q \bar{q} \rightarrow  s \bar{s}$ ,
$g \bar{g} \rightarrow  s \bar{s}$.
The cooling of the heat bath is governed by the following 
equation~\cite{ijmpa}:
\be
\frac{dT}{d\tau}=-c_s^2\frac{T}{\tau}
-\frac{b(\dot{r_s}+\dot{r_{\bar{s}}})}{\alpha(a+b(r_s+r_{\bar{s}}))}
\label{eq3}
\ee
where $\alpha=(1+c_s^2)/c_s^2$, $a=8\pi^2/(45c_s^2)$ and
$b=7\pi^2n_F/(120c_s^2)$,  $c_s^2$ is the velocity of sound, $n_F$
is the number of flavours and $\dot{r_s}=dr_s/d\tau$. In case
of (equilibrium) Bjorken's scaling solution~\cite{jdb} the last term in 
Eq.~\ref{eq3} vanishes. 

When the temperature of the QGP phase approaches  the transition 
temperature $T_c$ due to expansion, then the $s$ and $\bar{s}$ 
quarks hadronise to strange hadrons like $K^+$, $K^-$, $\Lambda$  
etc. An exhaustive set of hadronic reactions has been considered
for the production of $K^+$ in the hadronic 
phase~\cite{lileebrown} (see also~\cite{jajati}).

The evolution of $K^+$ ($u\bar{s}$) in the mixed phase is governed by
the following equation:

\bea
\frac{dr_{K^{+}}}{d\tau}&=&\frac{R_{K^{+}}(T_c)}{n^{eq}_{K^{+}}}
\left(1 - r_{K^{+}}r_{K^{-}}\right)
+\frac{R_{\Lambda}(T_c)}{n^{eq}_{K^{+}}}
\left(1-r_{K^{+}}r_{\Lambda}\right)\nn\\
&+&\frac{R_{\Sigma}(T_c)}{n^{eq}_{K^{+}}}
\left(1-r_{K^{+}}r_{\Sigma}\right)
+\frac{1}{f}\frac{df}{d\tau}\left(\delta\frac{r_{\bar{s}}n^{eq}_{\bar{s}}}
{n^{eq}_{K^{+}}}
- r_{K^{+}}\right),
\label{eq2}
\eea
Similar coupled equations can be written for $\Lambda$ and $\Sigma$. 
In the above equation $f$ represents  the fraction of hadrons 
in the mixed phase at time $\tau$.
The last term stands for the hadronization of
$\bar{s}$ quarks to $K^+$~\cite{kapusta,matsui}. Here $\delta$ is a
parameter which indicates the fraction of $\bar{s}$  quarks
hadronizing to $K^+$.The value of $\delta=0.5$
if we consider  only $K^+$ and $K^0$ formation
in the mixed phase.  
The initial values
of $\bar{s}$ quarks are taken close to their equilibrium values.
However, a small change in the initial value of $r_{\bar{s}}$
does not change the final results
significantly. Even with lower initial values of $r_{\bar{s}}$
the system reaches equilibrium very fast due to their production
in the high temperature heat bath.
To obtain the particle density we
solve the above-mentioned coupled set of differential equations 
numerically.

\begin{figure}[t]
\begin{center}
\includegraphics[width=7.0cm]{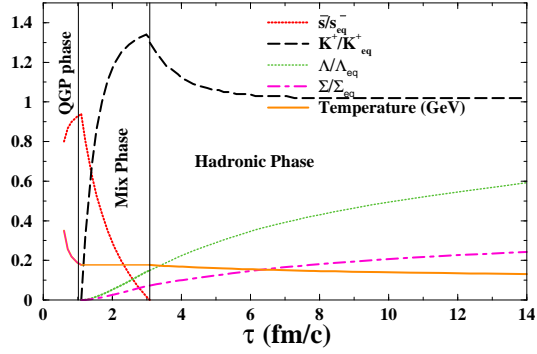}
\caption{%
The time evolution of the ratio of non-equilibrium to 
equilibrium density of quarks/various hadrons when initial
phase is assumed to be QGP ($T_i > T_c $ ).
Here $\bar{s}$, $K^+$ etc. stand for their corresponding  
number densities.
}
\label{fig1}
\end{center}
\end{figure}

\begin{figure}[b]
\centerline{
 \psfig{width=5.5cm,  figure=\pathnow 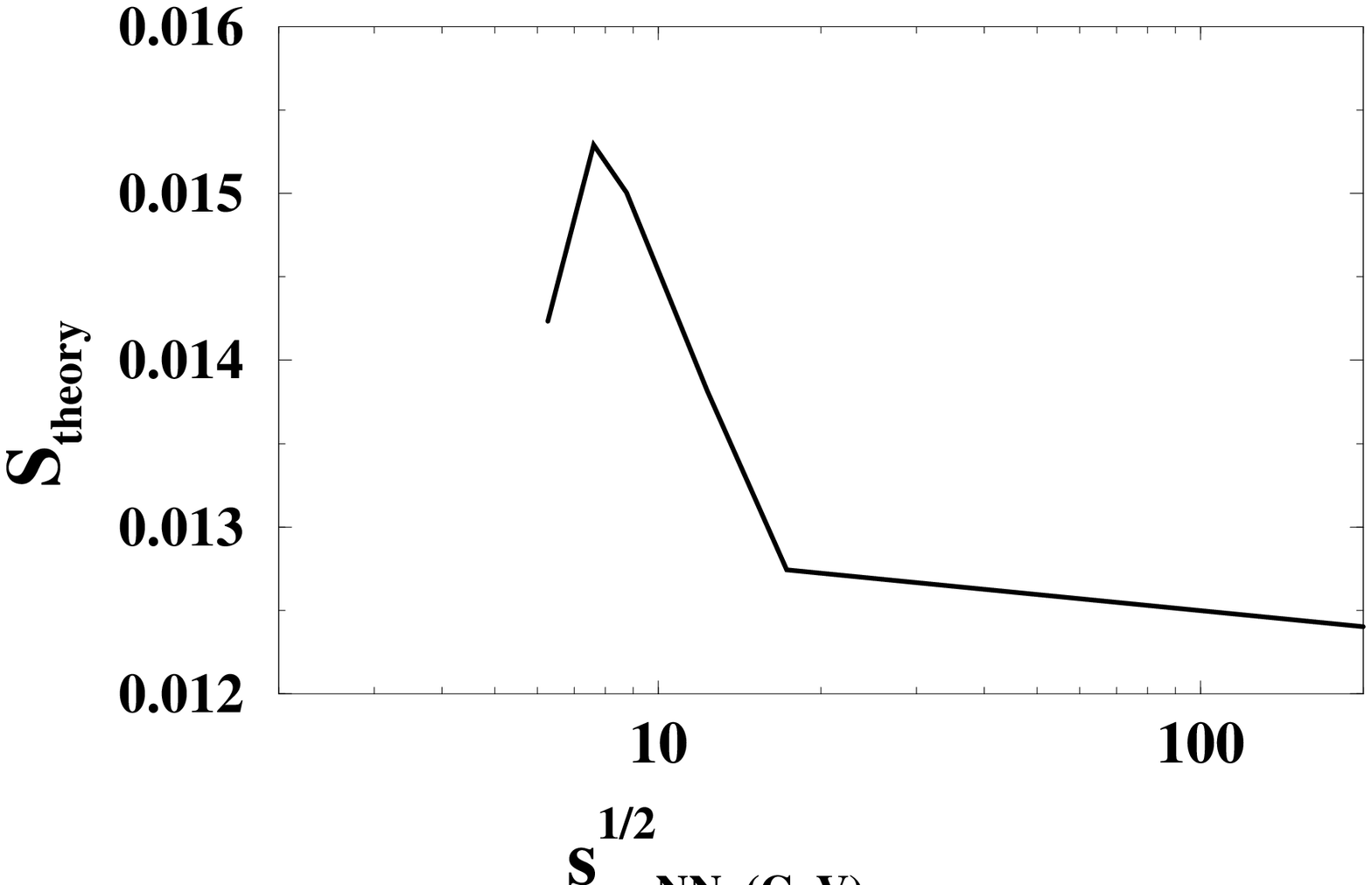}
\hspace*{-0.09cm}
\psfig{width=5.5cm, figure=\pathnow  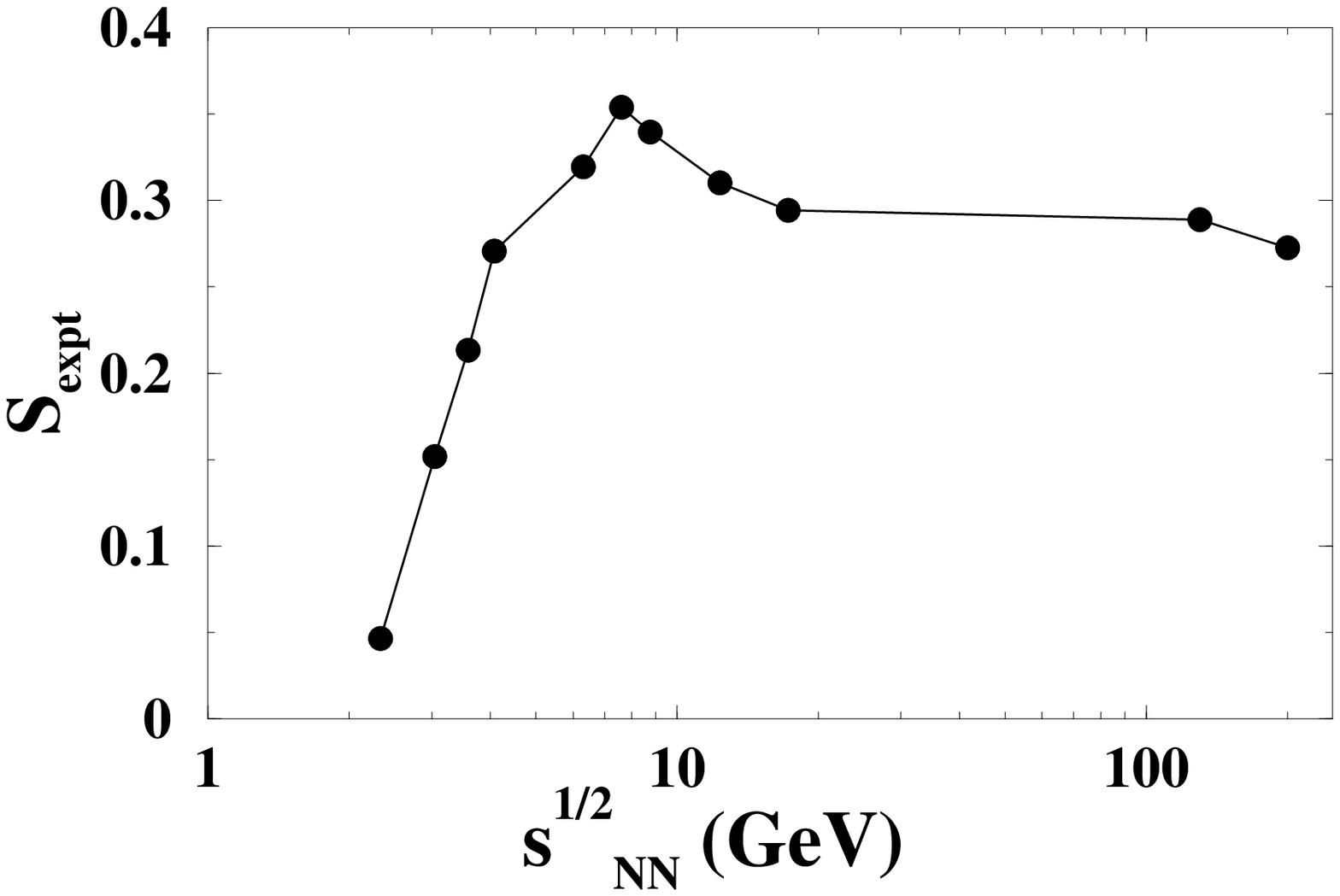}
}
  \caption{
The variation of anti-strangeness to entropy ratio, $S_{theory}$ (left) 
and $S_{expt}$ (right) with $\sqrt{s_{NN}}$  (see text).
}\vskip -.6cm
\label{fig2}
\end{figure}

The time evolutions of $r_{i}$ are shown in Fig.~\ref{fig1}.
We observe a clear over saturation in number density of kaons
at the end of the  mixed phase.
For $T_i=T_c$ or $T_i>T_c$, we get 
$r_{K^+}\sim $  1.2 - 1.4
at the end of the mixed phase, before it reaches the equilibrium value
of 1 in the hadronic phase.
For $T_i<T_c$ {\it i.e.} if the system is formed
in hadronic phase, we observe that $r_{K^+} < 1$,
for $\delta=0.5$. For smaller values of $\delta$, $r_{K^+}$ will
be even smaller. This indicates that the strangeness remains out of
chemical equilibrium if the system is formed in the hadronic phase. 
For a given centrality and 
$\sqrt{s_{NN}}$ (c.m.energy) we take particle multiplicity ($dN/d\eta$) 
and evaluate the initial temperature using.

\begin{equation}
T_i^3=\frac{2 \pi^4}{45 \zeta (3)}
         \frac{1}{\pi R^2 \tau _i}
         \frac{90}{4\pi^2g_{eff}}\frac{dN}{d\eta},
\end{equation}
where $\zeta(3)$ denotes the Riemann zeta function, $R$ is the 
transverse radius [$\sim$ $1.1(N_{part}/2)^{1/3}$,
$N_{part}$ is the number of participant nucleons ] of 
the colliding system , ${\tau}_i$ is the initial time and $g_{eff}$ 
is the statistical degeneracy.
For initial temperatures corresponding to each $dN/d\eta$ 
(hence $\sqrt{s_{NN}}$) we solve for 
$S_{theory}$ ($\equiv$ anti-strangeness/entropy) which is 
proportional to $K^+/\pi^+$.
In Fig.~\ref{fig2} (left) we depict the variation of 
$S_{theory}$ with $\sqrt{s_{NN}}$ which shows a non-monotonic (horn-like) behaviour.


\begin{figure}[t]
\begin{center}
\includegraphics[width=7.5cm]{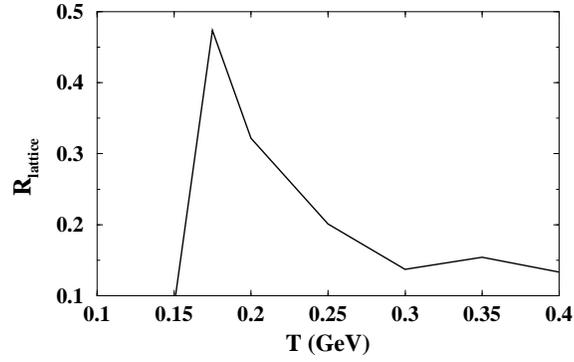}
\caption{Variation of $R_{lattice}$ with temperature. 
lattice QCD results taken from Ref.~\protect{\cite{lattice}}.
}
\label{fig3}
\end{center}
\end{figure}

Taking $dN_{K^+}/d\eta$, $dN_{K^-}/d\eta$ and $dN_{tot}/d\eta$
from experiments~\cite{MG,expt} we calculate the fractional entropy carried
by the strange sector, which will be proportional to $S_{expt}(=
g_{eff}^s/g_{eff}^{tot})$
where $g_{eff}^s$ ($g_{eff}^{tot}$) is the degrees of freedom associated 
with the strange (all) hadrons. 
In Fig.~\ref{fig2} (right)
we show the variation of $S_{expt}$ with $\sqrt{s_{NN}}$. Does lattice
QCD indicate any such behaviour?
To understand this we plot 
the quantity, 
${R_{lattice}}\equiv[{g^{(2+1)F}-g^{2F}}]/{g^{(2+1)F}}$ derived
from lattice QCD as a 
function of temperature ($T$) in Fig.~\ref{fig3}.
Here $g^{(2+1)F}$ and $g^{2F}$ are the effective statistical
degeneracies for (2+1) flavours and
2 flavours respectively. The quantity, $R_{lattice}$
 reflects the effective strange degrees of freedom relative to the total
{\it i.e.} it is the fractional entropy carried by the strange and 
anti-strange particles.  
 One clearly observes a  peak  near $T_c$ 
($\sim 175$ MeV) followed by a reduction in the values at higher
temperatures. This non-monotonic behaviour indicates the deconfinement
of the quarks and gluons.
\par
In summary, we have used the Boltzmann equation to study
the evolution of $\bar{s}$ and $K^+$ in heavy ion collision.
The horn-like structure observed in the $K^+/\pi^+$ 
arises due to the sudden increase of entropy resulting from the release
of large number of (colour) degrees of freedom at the deconfinement
transition point.
\par
\begin{ack}
Authors are  grateful to  B. Tomasik, M.Gazdzicki and N. Xu 
for useful discussions.
\end{ack}


\begin{thebibliography}{99}

\bibitem{lattice} 
\refer{F. Karsch}{ Nucl. Phys. A}{698}{2002}{199c}

\bibitem{npa757} `First Three Years of Operation of RHIC', Nucl.
Phys. A {\bf 757} (2005) 1-283 .

\bibitem{MG} M. Gazdzicki, J. Phys. G {\bf 30} (2004) S701.

\bibitem{CORS}  J. Cleymans, H. Oeschler, K. Redlich and 
S. Wheaton,  hep-ph/0510283. 
                         
\bibitem{GG} M. Gazdzicki and M. Gorenstein,  Acta Phys. Polon. B {\bf 30} 
(1999) 2705.

\bibitem{BT} B. Tomasik, nucl-th/0509101.

\bibitem{LR}  J. Letessier and J. Rafelski, nucl-th/0504028.

\bibitem{ijmpa} J. Alam, P. Roy, S. Sarkar, S. Raha and B. Sinha,
Int. J. Mod. Phys. A {\bf 12} (1997) 5151.

\bibitem{jdb} J. D. Bjorken, Phys. Rev. D {\bf 27} (1983) 140.

\bibitem{lileebrown} G. Q. Li, C. H. Lee and G. E. Brown, Nucl. Phys. A
{\bf 625} (1997) 372.

\bibitem{jajati} Jajati. K. Nayak, J. Alam, B. Mohanty, P. Roy and 
A. K. Dutt-Mazumder, nucl-th/0509061.

\bibitem{kapusta} J. Kapusta and A. Mekjian, Phys. Rev. D {\bf 33} (1986) 1304.

\bibitem{matsui} T. Matsui, B. Svetitsky, L. D. McLerran, Phys. Rev. 
D {\bf 34} (1986) 783; Phys. Rev. D {\bf 34} (1986) 2047.

\bibitem{expt} C. A. Ogilvie, {\it et al.}, Nucl. Phys. A {\bf 638} 57c (1998);
G. J. Odyniec, Nucl. Phys. A {\bf 638} 135c (1998); C. Blume, nucl-ex/0405010;
S. S. Adler {\it et al.}, Phys. Rev. C {\bf 69} 034909 (2004).

\end{thebibliography}
\end{document}